\documentclass[twocolumn]{aastex631}
\usepackage{amsmath}
\usepackage{amssymb}
\usepackage{graphicx}
\usepackage{enumitem}
\setcitestyle{notesep={; }}

\newcommand{\neowise}{{\it NEOWISE}}
\shorttitle{}
\shortauthors{}

\newcommand{\swift}{\textit{Swift}}
\newcommand{\nustar}{\textit{NuSTAR}}

\newcommand{\wtp}{WTP\,14adbjsh}

\begin{document}
	
	\title{A luminous dust-obscured Tidal Disruption Event candidate in a star forming galaxy at 42 Mpc}

	\correspondingauthor{Christos Panagiotou}
	\email{cpanag@mit.edu}
	
	\author{Christos Panagiotou}
	\affiliation{MIT Kavli Institute for Astrophysics and Space Research, Massachusetts Institute of Technology, Cambridge, MA 02139, USA}
	
	\author{Kishalay De}
	\altaffiliation{NASA Einstein Fellow}
	\affiliation{MIT Kavli Institute for Astrophysics and Space Research, Massachusetts Institute of Technology, Cambridge, MA 02139, USA}
	
	\author[0000-0003-4127-0739]{Megan Masterson}
	\affiliation{MIT Kavli Institute for Astrophysics and Space Research, Massachusetts Institute of Technology, Cambridge, MA 02139, USA}
	
	\author{Erin Kara}
	\affiliation{MIT Kavli Institute for Astrophysics and Space Research, Massachusetts Institute of Technology, Cambridge, MA 02139, USA}
	
	\author{Michael Calzadilla}
	\affiliation{MIT Kavli Institute for Astrophysics and Space Research, Massachusetts Institute of Technology, Cambridge, MA 02139, USA}
	
	\author[0000-0003-2895-6218]{Anna-Christina Eilers}
	\affiliation{MIT Kavli Institute for Astrophysics and Space Research, Massachusetts Institute of Technology, Cambridge, MA 02139, USA}
	
	\author{Danielle Frostig}
	\affiliation{MIT Kavli Institute for Astrophysics and Space Research, Massachusetts Institute of Technology, Cambridge, MA 02139, USA}
	
	\author{Viraj Karambelkar}
	\affiliation{Cahill Center for Astrophysics, California Institute of Technology, 1200 E. California Blvd. Pasadena, CA 91125, USA}
	
	\author{Mansi Kasliwal}
	\affiliation{Cahill Center for Astrophysics, California Institute of Technology, 1200 E. California Blvd. Pasadena, CA 91125, USA}
	
	\author{Nathan Lourie}
	\affiliation{MIT Kavli Institute for Astrophysics and Space Research, Massachusetts Institute of Technology, Cambridge, MA 02139, USA}
	
	\author{Aaron M. Meisner}
	\affiliation{NSF's National Optical-Infrared Astronomy Research Laboratory, 950 N. Cherry Ave., Tucson, AZ 85719, USA }
	
	\author{Robert A. Simcoe}
	\affiliation{MIT Kavli Institute for Astrophysics and Space Research, Massachusetts Institute of Technology, Cambridge, MA 02139, USA}
	
	\author[0000-0003-2434-0387]{Robert Stein}
	\affiliation{Cahill Center for Astrophysics, California Institute of Technology, 1200 E. California Blvd. Pasadena, CA 91125, USA}
	
	\author{Jeffry Zolkower}
	\affiliation{Cahill Center for Astrophysics, California Institute of Technology, 1200 E. California Blvd. Pasadena, CA 91125, USA}

	\begin{abstract}
		
		While the vast majority of Tidal Disruption Events (TDEs) has been identified by wide-field sky surveys in the optical and X-ray bands, recent studies indicate that a considerable fraction of TDEs may be dust obscured, and thus preferentially detected in the infrared (IR) wavebands. In this Letter, we present the discovery of a luminous mid-IR nuclear flare (termed \wtp) identified in a systematic transient search of archival images from the NEOWISE mid-IR survey. The source reached a peak luminosity of $L \simeq 10^{43} \text{erg s}^{-1}$ at 4.6 $\mu$m in 2015, before fading in the IR with a TDE-like $F \propto t^{-5/3}$ decline, radiating a total of more than $ 3\times 10^{51}$\,erg in the last $7$ years. The transient event took place in the nearby galaxy NGC 7392, at a distance of around $ 42$ Mpc; yet, no optical or X-ray flare is detected. We interpret the transient as the nearest TDE candidate detected in the last decade, which was missed at other wavelengths due to dust obscuration, hinting at the existence of TDEs that have been historically overlooked. Unlike most previously detected TDEs, the transient was discovered in a star forming galaxy, corroborating earlier suggestions that dust obscuration suppresses significantly the detection of TDEs in these environments. Our results demonstrate that the study of IR-detected TDEs is critical in order to obtain a complete understanding of the physics of TDEs, and to conclude whether TDEs occur preferentially in a particular class of galaxies.
		
	\end{abstract}
	
	
	
	\section{Introduction} 
	\label{sec:intro}
	
	A tidal disruption event (TDE) occurs when a star passes sufficiently close to a supermassive black hole so that tidal forces overcome the star's self-gravity, leading to its destruction. Part of the stellar debris accretes onto the black hole via a temporarily formed disk, which results in a transient flare of emission across the electromagnetic spectrum \citep[e.g.][]{1979SvAL....5...16L, 1988Natur.333..523R}. TDEs offer a unique opportunity to study the properties of quiescent supermassive black holes in the center of galaxies, which are otherwise unobservable. They also comprise an ideal laboratory to probe the physics of accretion onto black holes, such as the formation of an accretion disk and its evolution, on human time scales. 
	
	Given the typical mass of the central black holes ($M_\text{BH} \sim 10^{6}-10^8 M_\odot$)
	and assuming a luminosity close to the Eddington limit, the TDE spectrum is expected to peak in the soft X-rays. It is, thus, not a surprise that the first TDEs were discovered with ROSAT in the soft X-ray band \citep[e.g.][]{1996A&A...309L..35B, 1999A&A...349L..45K}. Nevertheless, the field of TDE studies was revolutionized by recent wide-field optical surveys, which, to date, have resulted in the detection of several dozens of TDE candidates, providing a wealth of information about the physics of these events \citep[e.g.][]{2014ApJ...793...38A, 2019MNRAS.488.1878N, 2019ApJ...873...92B, 2019ApJ...883..111H, 2021ApJ...908....4V, 2022arXiv220301461H}. In addition, the recent launch of \textit{SRG}/eROSITA \citep{2021A&A...647A...1P} has enabled the detection of more TDE candidates in the X-rays \citep{2021MNRAS.508.3820S}.
	
	As the number of known TDEs has been continuously increasing over the last years, it was noticed that the detected TDEs appear to be preferentially hosted in a specific class of galaxies \citep{2014ApJ...793...38A}. In particular, \cite{2016ApJ...818L..21F} noted that a large fraction of TDE hosts belong to an unusual group of quiescent galaxies with recently terminated star bursting activity. Later studies with a larger sample of sources concluded that the TDE hosts occupy predominantly a rare area of the galaxy parameter space between the blue star forming and the red quiescent galaxies, which is often referred to as the green valley \citep[e.g][]{2021ApJ...908....4V}. 
	
	In a recent exploration of potential selection effects in observing TDEs, \cite{2021ApJ...910...93R} demonstrated that even if TDEs occur commonly in star-forming galaxies, with a rate similar to the rate of detected TDEs, their detection may be significantly suppressed due to dust obscuration in the host galaxy. In this case, the obscured TDE may be detected in the infrared (IR) wave band, where dust thermal emission peaks. Interestingly, a few TDEs have already been detected serendipitously in IR surveys searching for dust obscured supernovae in peculiar galactic environments \citep{2018Sci...361..482M, 2020MNRAS.498.2167K}. In addition, a robust TDE candidate with a prominent IR echo was found in a nearby star forming galaxy by \cite{2022ApJ...930L...4W}. 
	
	Motivated by these results, we recently initiated a systematic search for IR luminous TDEs in archival images from the NEOWISE \citep{Mainzer2014} survey. A similar search of the NEOWISE data for extragalactic transients was recently performed by \cite{2021ApJS..252...32J} as well. Using the integrated NEOWISE photometry of galaxies with an estimated redshift in the Sloan Digital Sky Survey \citep[SDSS,][]{2002AJ....124.1810S}, these authors identified numerous mid-IR flares in low-redshift galaxies. Most of these flares were associated with the nucleus of the galaxy, pointing at an active galactic nucleus (AGN) or TDE nature for the transient, while only a fraction of these events had a corresponding optical flare, demonstrating that numerous nuclear transient events may be missed by optical surveys. In contrast to the aforementioned study, we identified mid-IR transients performing image subtraction on the public NEOWISE data (Sect. \ref{sec:detect}) in order to increase the sensitivity of our search. Furthermore, we did not restrict our sample only to SDSS galaxies. 
	
	In this Letter, we present the analysis of the brightest TDE candidate that resulted from our search. The initial identification and follow-up observations of the source are presented in Section \ref{sec:obs}. Section \ref{sec:analysis} presents the main results of our analysis, which are further discussed in Section \ref{sec:disc}. Throughout the rest of the manuscript we adopt a flat $\Lambda$CDM cosmology with $H_0 = 67.8 ~\text{km s}^{-1}~\text{Mpc}^{-1}$. All the reported errors correspond to 1$\sigma$ confidence interval, unless otherwise noted.

	
	\section{Observations} \label{sec:obs}
	\subsection{WISE Discovery}
	\label{sec:detect}

	\begin{figure*}[!ht]
		\centering
		\includegraphics[width=\textwidth]{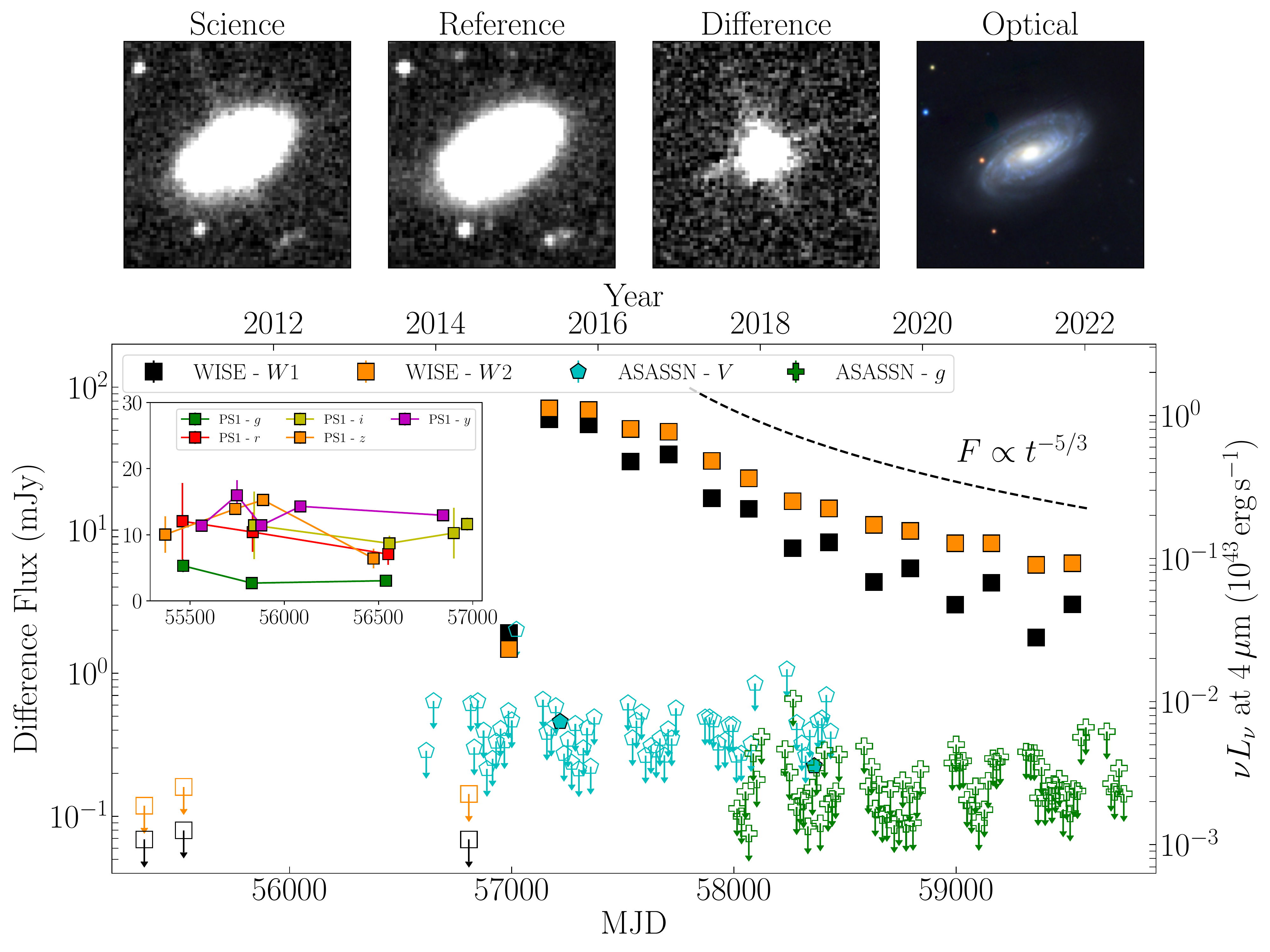}
		\caption{{\bf Multi-color light curves of WTP\,14adbjsh.} \textit{Upper panels}: Image cutouts at the location of NGC\,7392. The first three panels show the single epoch \neowise\ image at the time of the observed peak of the IR flare in 2015, the unWISE reference image created from stacking AllWISE data from 2010-11 and the difference image between the science and reference image respectively. The fourth panel shows an archival optical composite $gri$ image from PanSTARRS1. \textit{Lower panel}: Combined WISE (mid-IR) and ASASSN (optical) host subtracted light curves of the flare. The error bars are not distinguishable due to their nominal small value. The left $y$-axis denotes the observed flux of the transient, while the right $y$-axis denotes the corresponding monochromatic luminosity for a wavelength of $4\,\mu m$. The ASASSN data were stacked in 15-day bins to improve the SNR while the mid-IR light curve points correspond to individual visits from WISE separated by six months. The inset shows the optical PSF-flux measurements (without host subtraction) at the location of the nucleus from the PS1 catalog, demonstrating no evidence for an optical flare in the $5$\,years leading up to the mid-IR transient. }
		\label{fig:disc_lc}
	\end{figure*}

	The Wide-field Infrared Survey Explorer (WISE) satellite \citep{Wright2010}, re-initiated as the NEOWISE mission, has been carrying out an all-sky mid-infrared (mid-IR) survey in the $W1$ ($3.4$\,$\mu$m) and $W2$ ($4.6$\,$\mu$m) bands since 2014. As part of an ongoing program to identify large amplitude mid-IR transients in NEOWISE data, we carried out a systematic search for transients in time-resolved coadded images created as part of the unWISE project \citep{Lang2014, Meisner2018}. The details of this search will be presented in De et al. (in prep). In brief, we used a customized code \citep{De2019} based on the ZOGY algorithm \citep{Zackay2016} to perform image subtraction on the NEOWISE images using the full-depth co-added images of the WISE mission (obtained during 2010-2011) as reference images. 
	
	Cross-matching the sample of WISE transients to known nearby galaxies within 200\,Mpc \citep{Cook2019}, we identified the source WTP\,14adbjsh\footnote{For all transients identified in the WISE Transient Pipeline (WTP, De in prep.), we adopt this naming scheme indicating the year of the first detection followed a six letter alphabetical code. Further details will be presented in De et al. in prep.} as a bright mid-IR transient coincident with the nucleus of the galaxy NGC\,7392 (Figure \ref{fig:disc_lc}), at a redshift of $z = 0.0106$ \citep{Springob2005}. We adopt the mean and standard deviation of the detected positions in the difference images as our best estimate and its uncertainty for the transient position. The resulting positional measurements are $\alpha =$ 22:51:48.76, $\delta = -$20:36:28.98, with an uncertainty of $\approx 0.7$\arcsec, and it coincides with the reported nucleus of NGC 7392 within 0.3\arcsec. 
	
	To obtain the complete mid-IR light curve of the transient, we performed forced point spread function (PSF) photometry at the measured position of the transient on the WISE difference images. We converted the flux measured on the unWISE images to physical fluxes using the published WISE photometric zero-points\footnote{\url{https://wise2.ipac.caltech.edu/docs/release/allsky/expsup/sec4_4h.html}}. The resulting light curve, plotted in Fig. \ref{fig:disc_lc}, shows that the transient was first detected in the second epoch of NEOWISE observations in 2014, when it brightened by more than 3 orders of magnitude to $\sim 80$\,mJy ($\sim 8.5$\,mag) in the $W1$ and $W2$ bands in 2015. The source has been constantly fading since then but continues to be detected until the latest NEOWISE data release (2021) at a flux level of around $ 5$\,mJy in the $W2$ band. Imprinted on the slow fading, the light curve shows low level fluctuations on alternating epochs. Closer investigation of this trend reveals that they arise due to low-level uncorrected and systematic PSF variations along the two different scan directions in the successive WISE scanning epochs, and are not physically related to the transient. For a luminosity distance $D_\text{L} = 42.3 $ Mpc\footnote{Value retrieved from NED (\url{https://ned.ipac.caltech.edu/}).}, the MIR brightness of the flare corresponds to a peak $W2$ luminosity of $\nu L_\nu \simeq 10^{43}$\,erg\,s$^{-1}$.

	
	\subsection{Archival coverage}
	
	NGC\,7392 was observed in the optical bands as part of the ongoing All-sky Automated Survey for Supernovae (ASASSN; \citealt{Shappee2014}). We retrieved forced difference photometry light curves at the position of \wtp\ using the ASASSN Sky Patrol service \citep{Kochanek2017}. The transient is not formally detected in the single epoch photometry in $V$ and $g$ bands. We stack the photometry in bins of $\simeq 15$ days to obtain deeper limits given the long duration of the mid-IR transient. The mid-IR transient is not detected in the ASASSN optical bands to a flux limit more than 100 times deeper than the IR flux at peak (Fig. \ref{fig:disc_lc}).

	We also searched publicly available data from the Catalina Real-time Transient Survey (CRTS; \citealt{Drake2009}) for coverage of the source. While the source location was covered from around $2500$\,days to nearly $500$\,days prior to the first WISE detection, the nucleus is marked as saturated in the CRTS images and therefore not useful for analysis. We retrieved multi-band forced PSF photometry measurements at the position of the nucleus from the PanSTARRS1 survey \citep{Chambers2016} between the years 2010-2015. As the PS1 observations consist of several consecutive visits (typically within $10$\,days) spaced by several years in each filter, we compute the average flux of the nucleus over the closely spaced visits and adopt the scatter in the measurements as the uncertainty. We show the corresponding light curves in Fig. \ref{fig:disc_lc}, showing that the galaxy nucleus exhibits no evidence of an optical flare in the 5 years leading up to the mid-IR transient.
	As will be discussed in more detail below, the lack of an optical flare implies a significantly large absorption along our line of sight or an intrinsically faint source in the optical band.
	
	To search for transient X-ray emission, we used the online X-ray light curve generator of the Monitor of All-sky X-ray Image \citep[MAXI,][]{2009PASJ...61..999M} Gas Slit Camera\footnote{\url{http://maxi.riken.jp/mxondem/}}. We generated the light curve over the full history of the mission binned into intervals of 30 days. No significant emission is detected between 2010 and 2022 to a $5\sigma$ flux limit of $7.5 \cdot 10^{-3}$\,counts\,cm$^{-2}$\,s$^{-1}$. Assuming a power law X-ray spectrum with a nominal photon index of $\Gamma = 2$, we use the \texttt{WebPIMMS} calculator\footnote{\url{https://heasarc.gsfc.nasa.gov/cgi-bin/Tools/w3pimms/w3pimms.pl}} to derive a corresponding $2-6$\,keV flux limit of $F_{2-6 \text{keV}} < 4.5 \cdot 10^{-11}$\,erg\,cm$^{-2}$\,s$^{-1}$. If we instead assume a softer spectrum with $\Gamma = 4$ as observed in most X-ray TDEs \citep{2021MNRAS.508.3820S}, we constrain the X-ray flux to \mbox{$ F_\text{2-6 \text{keV}}< 5.6 \cdot 10^{-11}$\,erg\,cm$^{-2}$\,s$^{-1}$.} At the distance of NGC 7392, this flux limit corresponds to a luminosity of $L_\text{X} \lesssim 1.2 \cdot 10^{43} $\,erg\,s$^{-1}$.
	
	Finally, we retrieved publicly available multi-epoch quicklook images from the ongoing Very Large Array Sky Survey (VLASS; \citealt{Lacy2020}) at $2-4$\,GHz to search for radio emission at the transient position. The source has thus far been observed at two epochs in 2019 June and 2022 February (corresponding roughly to $1600$ and $2600$\,days after the first detection of the mid-IR flare, respectively). We created $30$\arcsec cutouts at the source position and estimated the peak flux and the standard deviation in the cutout maps to search for a point source. No source is detected in the region down to a $5\sigma$ threshold flux of around $0.75$\,mJy in both epochs.

	
	\subsection{Ground-based follow-up imaging}
	
	On October 19, 2022, we obtained one epoch of optical imaging of NGC\,7392 in the $ugri$ filters using the SUMMER camera mounted on the newly commissioned 40-inch telescope for the Wide-field Infrared Transient Explorer (WINTER; \citealt{Lourie2020}) at Palomar Observatory. We obtained single exposures of 300\,s in each filter, and the data were reduced using a custom pipeline for detrending, astrometry and photometric calibration. We used the PS1 catalog \citep{Chambers2016} for photometric calibration in the $gri$ filters and the SkyMapper catalog \citep{Onken2019} for the $u$-band calibration. We also obtained one epoch of multi-color near-infrared (NIR) imaging of NGC\,7392 on August 4, 2022, using the Spartan camera \citep{Loh2012} on the 4.1 m Southern Astrophysical Research (SOAR) Telescope as part of program SOAR 2022B-005 (PI: De). We obtained a series of 9 dithered exposures of the field with exposure times of 90\,s and 60\,s in the $J$ and $Ks$ filters respectively. The data were detrended followed by astrometric and photometric calibration against nearby 2MASS sources using a modified version of the imaging pipeline described in \citet{De2019}.

	
	\subsection{Ground-based spectroscopy}
	
	We obtained follow-up optical and near-infrared spectroscopy of the nucleus of NGC\,7392 to search for
	the presence of spectroscopic features associated with the nuclear flare. On July 4, 2022, we obtained a near-IR ($\approx 1.0 - 2.5\,\mu$m) spectrum using the Folded Port Infrared Echellette (FIRE; \citealt{Simcoe2013}) on the 6.5\,m Magellan Baade Telescope. Observations were acquired in the Echelle mode ($R \approx 5000$) using the 0.6\,\arcsec slit centered at the position of the nucleus of the host galaxy, for a total exposure time of 1800\,s on source in addition to sky exposures obtained by nodding the slit outside the galaxy light profile. The data were reduced and flux calibrated using observations of a nearby telluric standard using the \texttt{pypeit} \citep{Prochaska2020} package. 
	
\begin{figure*}
		\centering
		\includegraphics[width=\textwidth]{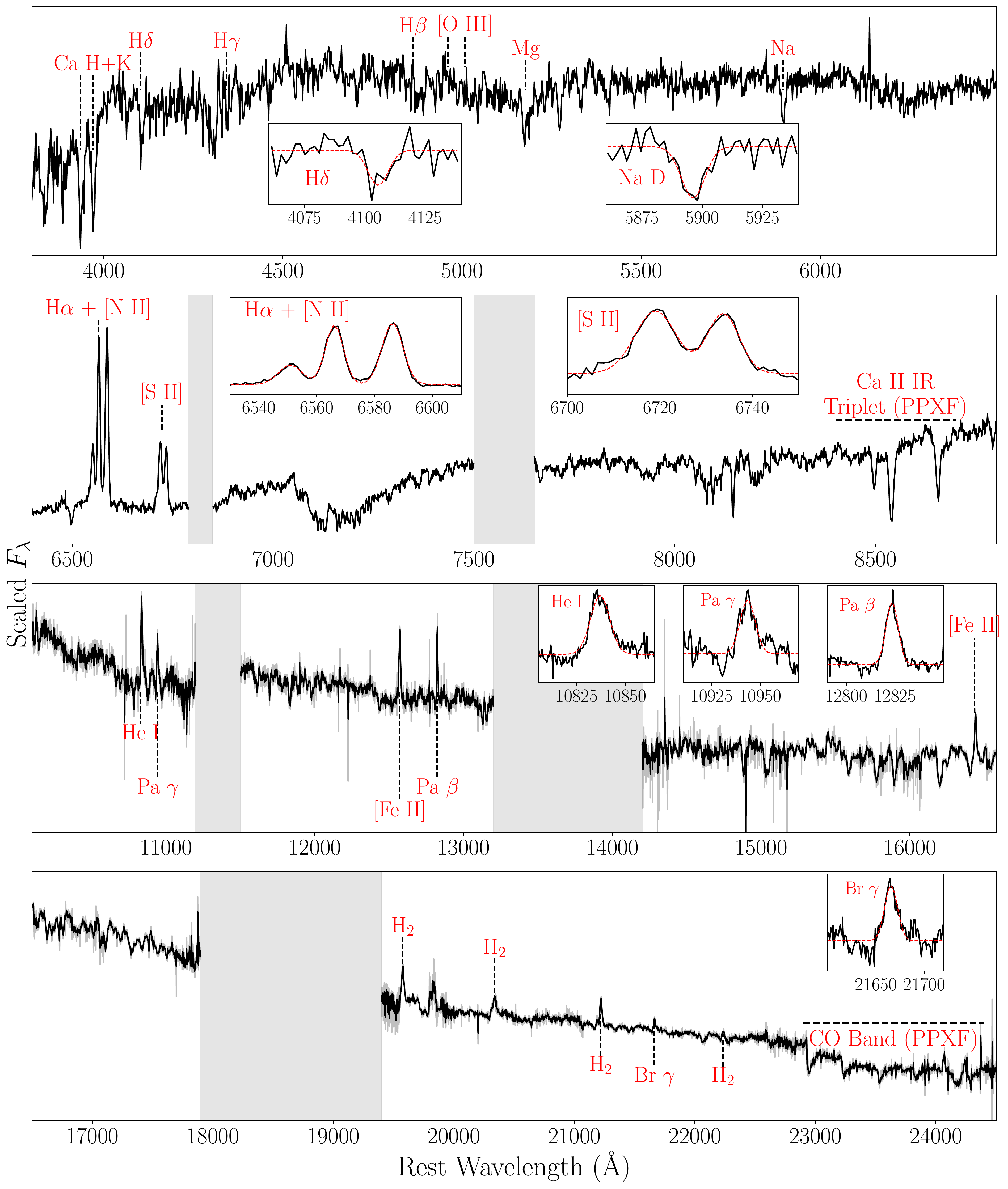}
		\caption{Optical and NIR follow-up spectroscopy of the nucleus of NGC\,7392. The gray lines show the raw spectra in the lower two panels while the black lines show the binned spectrum. Prominent spectral features are marked. Regions of telluric absorption and low atmospheric transmission are masked with gray boxes. The insets show profiles of the major emission and absorption features of H, He, N and Na together with best-fit Gaussian profiles. The spectral regions used for the PPXF fitting of the Ca II IR triplet and CO band-heads are indicated.}
		\label{fig:spec}
	\end{figure*}

	Further, we obtained a complete optical spectrum of NGC\,7392 on November 11 and 14, 2022, using the Goodman High Throughput Spectrograph \citep{Clemens2004} on SOAR as part of program SOAR 2022B-005 (PI: De). The data were acquired in the 600 lines/mm grating for the red side spectrum (630 - 893 nm; $R \simeq 1400$) and in the 400 lines/mm grating for the blue side spectrum (350 - 700 nm; $R \simeq 900$), for a total exposure time of 1200\,s and 600\,s respectively. Data reduction and flux calibration with a spectrophotometric standard was performed using again the \texttt{pypeit} package. The NIR and optical spectra are presented in Figure \ref{fig:spec}. 
		
	\subsection{X-ray observations}
	
	Following the IR discovery of \wtp, we obtained a \nustar\ and a \swift\ observation of the source through the approval of Director's Discretionary Time (DDT) requests. NGC 7392 was observed by the \swift\ observatory \citep{2004ApJ...611.1005G} on June 15, 2022 (ObsID 00015214001). We only consider the \swift\ XRT \citep{2005SSRv..120..165B} observation, which was performed in photon counting mode for an exposure time of 1.6 ks. The \nustar\ \citep{2013ApJ...770..103H} observation took place on July 12, 2022, with a net exposure of around 48.1 ks. Standardized procedure was followed to reduce the observational data using the HEASOFT\footnote{\url{https://heasarc.gsfc.nasa.gov/docs/software/lheasoft/}} software package. A circular region centered on the celestial coordinates of NGC 7392 was used to extract the source count rate. A radius of 20 (30) arcseconds was used for the \swift\ (\nustar) observation, while the background emission was extracted from a source free annular region encircling the source region. 
	
	The source is not significantly detected in X-rays (i.e. at more than 3$\sigma$ significance). The 3-78 keV average count rate of the \nustar\ observation is $0.0070 \pm 0.0034$ cnt s$^{-1}$, while the 0.3-10 keV \swift\ count rate is estimated to be $0.0024 \pm 0.0012$ cnt s$^{-1}$. Assuming a power-law spectrum with a photon index of $\Gamma = 2$, the above count rates indicate an upper limit on the X-ray source luminosity of $L_{2-10 \text{keV}} \lesssim 10^{40} ~\text{erg s}^{-1}$. The lack of a clear X-ray detection suggests that either the source is not currently bright in X-rays or potentially the source is heavily obscured by matter along our line of sight.

	
	\section{Results} 
	\label{sec:analysis}
	
	\subsection{IR flare evolution}
	\label{sec:flare}
	
	\begin{figure}
		\centering
		\includegraphics[width=\columnwidth, height=1.4\columnwidth, trim={10 0 10 0}, clip]{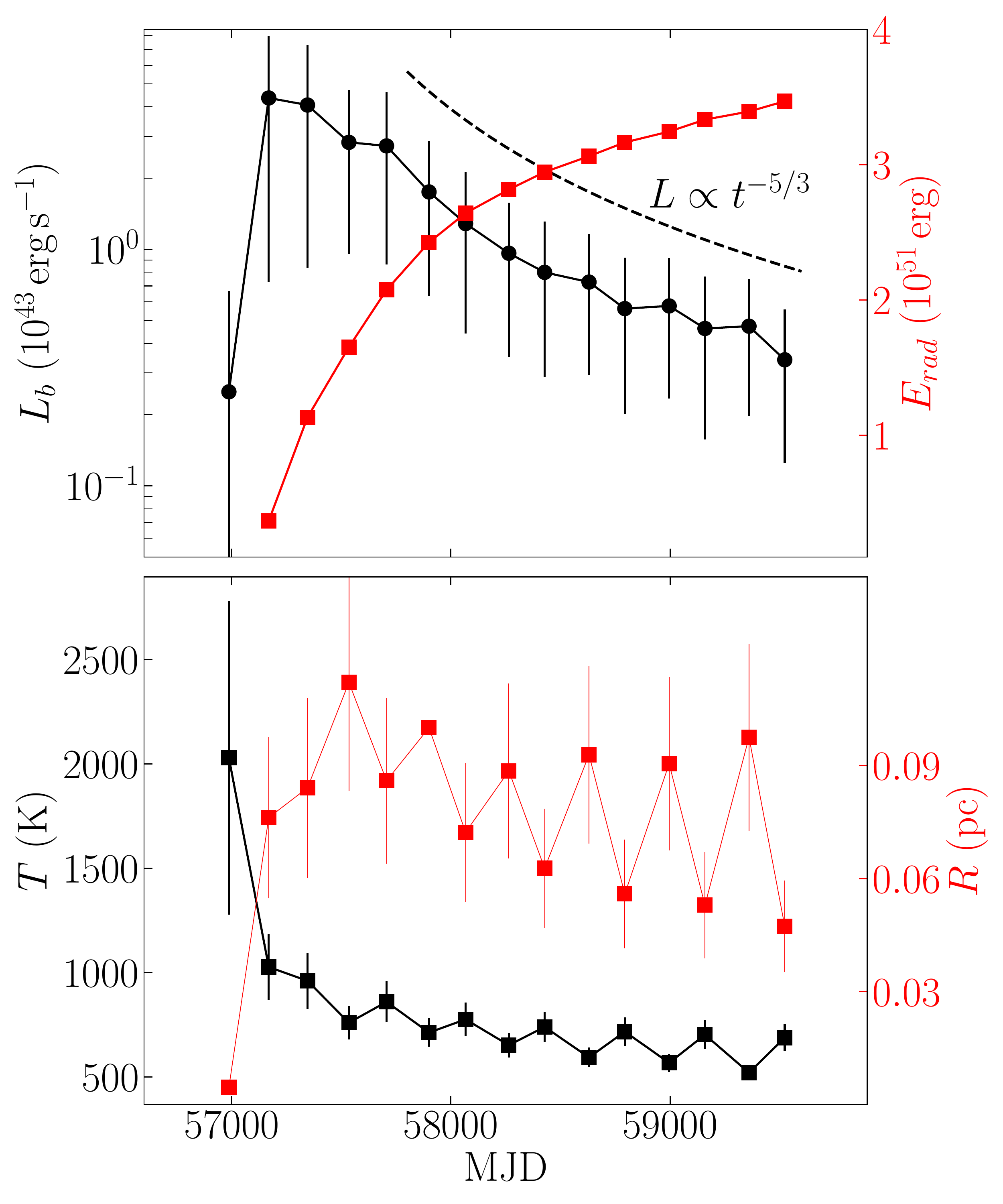}
		\caption{\textit{Upper panel}: Mid-IR bolometric luminosity (left axis) and total radiated energy (right axis) of \wtp\ as a function of time, overplotted with a $t^{-5/3}$ time evolution dependence for the luminosity. \textit{Lower panel}: The mid-IR color temperature (left axis) and effective radius (right axis) evolution of the transient. The low-level radius and temperature fluctuations in successive epochs during the post-peak fading are a result of the oscillating $W1$ flux (Figure \ref{fig:disc_lc}) that are caused by uncorrected and systematic PSF-shape variations on this very bright transient (Section \ref{sec:detect}).}
		\label{fig:bololc}
	\end{figure}

	We calculate the observed energetics of the mid-IR flare by fitting the $W1$ and $W2$ photometry with a simple black body function to derive the mid-IR luminosity and color temperature at each epoch. It should be mentioned that the dust emission is expected to follow a modified black body distribution, frequently called the gray body, which also depends on the dust composition \citep[e.g.][]{1982ApJ...252..589S, 1983QJRAS..24..267H}. However, an advanced modelling of the dust emission is outside the scope of the present work, especially since we have measurements in only two wave bands. We expect our best-fit results to be representative of the true physical properties, while our qualitative results should hold regardless of the exact dust emission model.
	
	In order to account for the systematic variations due to the changing PSF of the telescope in successive scans, we conservatively adopt a systematic uncertainty of 10\% in the flux measurements for the fitting procedure. We assume a flat prior on the emitting radius and temperature and use the Markov Chain Monte Carlo (MCMC) fitting library \texttt{emcee} \citep{Foreman-Mackey2013} to derive the best-fit luminosity and temperature (median of the posterior distributions), and their uncertainty intervals (inter-quartile range of the posterior distributions). The resulting evolution is shown in Fig. \ref{fig:bololc}. 
	
	The flare rises to a peak luminosity of around \mbox{$ 5\times 10^{43}$\,erg\,s$^{-1}$} in the first nearly $ 180$\,days followed by a slow decline over the subsequent years. The luminosity temporal evolution follows closely the expected evolution for the mass fallback rate of a TDE \citep[e.g.][dotted line in Fig. \ref{fig:bololc}]{1989IAUS..136..543P}. The temperature evolution shows a rapid cooling in the first two epochs of detection from around $2000$\,K  (albeit consistent with the dust sublimation temperature within the errors) to less than $ 1000$\,K, followed by a slow cooling in the next epochs. Except for the first epoch, the temperature is consistently lower than 1500 K and thus, the emitting dust is consistent with both a graphite and silicate grain composition, which have a sublimation temperature of 1800 and 1500 K, respectively. 
	
	The upper panel of Fig. \ref{fig:bololc} shows the large cumulative energy output of the transient source, observed in the mid-IR, $E_\text{rad, BB} > 3 \cdot 10^{51}~\text{erg}$, which was estimated using the best-fit values of the bolometric luminosity. This number should be interpreted as a strict lower limit of the total energy output of the source, because it is estimated using only the observed mid-IR emission, which corresponds to the dust reprocessed emission of the source intrinsic flux, and because we assumed a black body law for the dust emission, instead of a less efficient gray body spectrum. If the dust covering of the source is considerably lower than 1 or if the source is bright in wavelengths, such as the hard X-rays, that are not heavily obscured by a dusty medium, the total energy output of this transient would be significantly larger than $E_\text{rad, BB} $.
	
	Moreover, since the detected emission most likely corresponds to the UV reprocessing of a central source by a surrounding dusty medium, the shape of the IR light curve may be used to deduce the physical properties of that medium \citep{2016ApJ...828L..14J, 2016ApJ...829...19V}. For instance, assuming an isotropically illuminating source and a spherical distribution for the dusty medium, the rapid rise of the IR light curve and the temporal duration of the observed maximum luminosity constrain the maximum diameter of the dusty sphere to $d_\text{max} \leq 1~\text{ly} \simeq 0.31~\text{pc}$ from the inner source. The rather small spatial extent of the reprocessing dusty medium (relative to the WISE cadence) suggests that the WISE light curve approximately reflects the shape of the driving light curve of the inner source, scaled to the appropriate amplitude. In other words, because the reprocessing medium is small, we expect that to first order the intrinsic (unobserved) UV emission has a similar temporal evolution to the mid-IR light curve. 
		
	A minimum value on the inner radius of the dusty medium may be obtained by estimating the dust sublimation radius. Following \cite{2008ApJ...685..160N}, the sublimation radius is given by:
	
	\begin{equation}
		R_\text{sub} = 0.4 \left ( \frac{L}{10^{45} ~\text{erg s}^{-1}} \right )^{0.5} \left ( \frac{1500 K}{T_\text{sub}} \right )^{2.6} \mathrm{pc},
	\end{equation}

	\noindent where L is the illuminating luminosity and $T_\text{sub}$ is the sublimation temperature of the dust grains. For the typical graphite sublimation temperature, $T_\text{sub} \simeq 1800 K$, and $L = 5 \cdot 10^{43} ~\text{erg s}^{-1}$, we obtain $R_\text{sub} \simeq 0.056 ~\text{pc}$, which is comparable to the estimated black body emission radius (Fig. \ref{fig:bololc}).

	
	\subsection{Black hole mass estimate}
	
	We use the high resolution NIR spectrum of the nucleus to estimate the mass of the supermassive black hole in the nucleus of NGC\,7392. We use the Penalized Pixel Fitting (\texttt{PPXF}) routine \citep{Cappellari2022} to measure the line-of-sight velocity dispersion using the absorption lines in the spectra. We fit the wavelength region of $22900 - 24000$\,\AA\ which contains the CO absorption bandheads (Fig. \ref{fig:spec}). We use the GNIRS spectral template library \citep{Winge2009} as the reference library, which contains spectra of 60 late-type stars. The velocity dispersion is estimated by convolving the template stellar spectra with the line-of-sight velocity dispersion. The template spectra were first degraded to the resolution of the observed spectra, followed by fitting using the default options for the software. The best-fit model provides a velocity dispersion of $\sigma_{\rm CO} = 117 \pm 7$\,km\,s$^{-1}$. Performing the same measurement using the Ca\,II triplet with the spectral libraries in \citet{Cenarro2001}, we measure a higher value of $\sigma_{\rm Ca} = 143 \pm 8$\,km\,s$^{-1}$, consistent with previously reported biases between the two dispersion indicators \citep{Riffel2015}. Using the scaling relationship between the black hole mass and bulge dispersion velocity in \citet{Kormendy2013}, we derive $\log (\frac{M_{BH}}{M_\odot}) = 7.5^{+0.2}_{-0.3}$ from the CO lines and $\log (\frac{M_{BH}}{M_\odot}) = 7.8^{+0.2}_{-0.2}$ from the Ca\,II lines. The two measurements are consistent within error bars, noting that the relationship has an intrinsic scatter of $0.29 \pm 0.03$.

	
	\subsection{Host galaxy properties}

	\subsubsection{Integrated photometry}
	\label{sect:host_photo}

	We perform photometry on the multi-band optical ($ugri$) and NIR images obtained as part of our follow-up observations, in order to model the integrated properties of NGC\,7392. We used the python package \texttt{petrofit} \citep{Geda2022} to perform source detection on the pixel-level data followed by deblending of nearby stars and calculation of global shape parameters. We then calculate fluxes contained within concentric elliptical annuli matched to the shape of the galaxy to carry out a curve-of-growth analysis for deriving the total flux of the galaxy stellar light in the respective filters. We additionally retrieved pre-flare integrated photometry for the host galaxy in the ultraviolet filters from the Galaxy Evolution Explorer (GALEX; \citealt{Iglesias2006}) source catalog and in the mid-IR filters from the AllWISE catalog (prior to the start of the nuclear flare; \citealt{Cutri2013}). The compiled photometry measurements are provided in Table \ref{tab:host_photo}.

	
	\begin{table}[!ht]
		\centering
		\caption{Measured integrated host magnitudes for NGC\,7392 in the optical and NIR bands (corrected for Galactic extinction). While the optical/UV magnitudes were directly calibrated to the AB magnitude system, the NIR 2MASS (Vega) magnitudes were converted to AB mags assuming corrections of $0.91$ and $1.85$ mags in the $J$ and $Ks$ filters respectively. For the WISE filters, we used the published photometric zero-point fluxes to convert Vega magnitudes to the AB system.}
		\begin{tabular}{ccc}
			\hline
			\hline
			Filter & Wavelength ($\mu$m) & AB mag  \\
			\hline
			GALEX $1530$ & $0.153$ & $14.87 \pm 0.02$\\
			GALEX $2315$ & $0.231$ & $14.53 \pm 0.01$\\
			$u$ & $0.354$ & $13.48 \pm 0.04$\\
			$g$ & $0.477$ & $12.18 \pm 0.02$\\
			$r$ & $0.623$ & $11.57 \pm 0.01$\\
			$i$ & $0.762$ & $11.23 \pm 0.01$\\
			$J$ & $1.235$ & $10.55 \pm 0.01$\\
			$Ks$ & $2.159$ & $10.64 \pm 0.01$\\
			WISE $W1$ & $3.352$ & $11.74 \pm 0.01$ \\
			WISE $W2$ & $4.603$ & $12.35 \pm 0.01$ \\
			WISE $W3$ & $11.561$ & $11.34 \pm 0.01$ \\
			WISE $W4$ & $22.088$ & $11.16 \pm 0.01$ \\
			\hline
		\end{tabular}
		
		\label{tab:host_photo}
	\end{table}
	

	
	\begin{figure}
		\centering
		\includegraphics[width=0.52\textwidth, height=0.6\textwidth, trim={0 30 0 50}, clip]{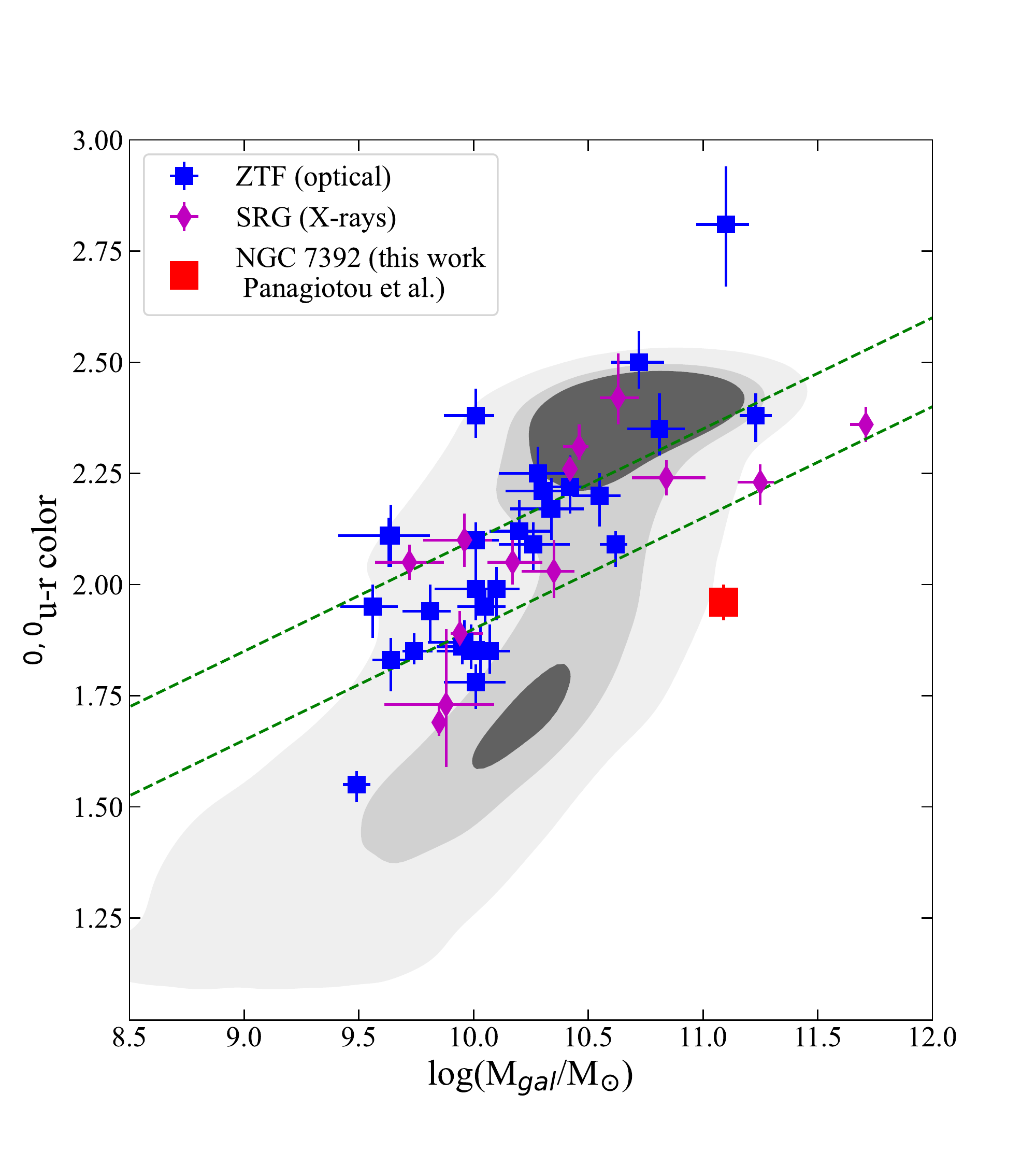}
		\caption{The Galactic extinction corrected, rest frame \mbox{$u-r$} color of NGC 7392 (red square) versus its best-fit stellar mass, obtained through reproducing the broadband photometry of the galaxy. The errors are not distinguishable due to their small value. Magenta diamonds denote the values for TDEs discovered by \textit{SRG}/eROSITA \citep{2021MNRAS.508.3820S}, while blue squares correspond to TDEs found in the ZTF survey \citep{2022arXiv220301461H}. The contours enclose a comparison sample of nearby ($z<0.1$) SDSS galaxies \citep{2014ApJS..210....3M}. The green dashed lines denote the limits of the green valley parameter space, as given by \cite{2021ApJ...908....4V}, where most of the optically and X-ray selected TDEs are found. Contrarily, NGC 7392 sits in a very different region of parameter space, where star-forming galaxies reside.  } 
		\label{fig:urmgal}
	\end{figure}	

	In order to compare the host galaxy of \wtp\ to those of typical optical and X-ray detected TDEs, we estimated the physical properties of NGC 7392 following the procedure outlined by \cite{2014ApJS..210....3M} and \cite{2021ApJ...908....4V}. In brief, we use the Prospector software \citep{2021ApJS..254...22J}, which employs a flexible stellar population synthesis
	module \citep[FSPS,][]{2009ApJ...699..486C, 2010ApJ...712..833C} to reproduce photometric data. We use the same model considered by previous TDE studies \citep[e.g.][]{2021ApJ...908....4V}, which assumes a smoothly declining star formation history. This model has only five free parameters, the stellar mass, $M_{\rm gal}$, the metallicity, the dust optical depth \citep[using the extinction law of][]{2000ApJ...533..682C}, the age of the stellar population, and the e-folding timescale characterizing the decrease of star formation. We obtained the statistical uncertainties of the best-fit parameters through MCMC sampling \citep{Foreman-Mackey2013}, and we used the best-fit results to estimate the Galactic extinction corrected, rest frame $u-r$ color. We found that $\log (\frac{M_{\rm gal}}{M_\odot}) = 11.08 \pm 0.01 $ and $^{0,0}u-r = 1.96 \pm 0.04 $.
	
	Figure \ref{fig:urmgal} plots the $u-r$ color versus the stellar mass for NGC 7392 and for the latest sample of TDE candidates from the ZTF survey \citep{2022arXiv220301461H} and \textit{SRG}/eROSITA \citep{2021MNRAS.508.3820S}. It is interesting to note that while optically and X-ray detected TDEs are found predominantly in galaxies in the green valley parameter space, NGC 7392 lies in the star forming region of the plot, corroborating previous suggestions that optical and X-ray surveys may provide a biased view of the TDE population. This point is further discussed in Sect. \ref{sec:disc}.

	
	\subsubsection{Spectroscopic line diagnostics}
	We use the prominent emission and absorption lines in the optical spectra to obtain independent estimates of the host galaxy type. We fit the region around the H$\alpha$ + [N II] complex using a sum of three Gaussian functions (shown in Figure \ref{fig:spec}) to estimate the line fluxes, velocity dispersions, and equivalent widths. We also similarly fit the H$\delta$ and Na D absorption lines in the blue side spectrum to obtain the corresponding equivalent widths. The blue side spectrum shows clear evidence for weak emission in the H$\beta$ waveband and a less prominent narrow \lbrack O III\rbrack\ line. In order to estimate the line fluxes, we first subtract the stellar absorption features by fitting them with the \texttt{PPXF} routine using spectral templates in the MILES library \citep{Falcon-Barroso2011}, followed by performing the same Gaussian fitting on the residual emission lines. We estimate uncertainties on the corresponding measurements by adding simulated noise to the spectrum scaled to the root-mean-square of the residual fluctuations from the best-fit model. The resulting measurements are provided in Table \ref{tab:host_spec}.
	
	\begin{table*}[]
		\centering
		\caption{Spectroscopic line flux, velocity dispersions and equivalent widths in the spectrum of the nucleus of NGC\,7392. The dagger symbol $^\dagger$ denotes measurements from absorption features. We caution that the blue side ($\lambda < 6000$\,\AA) and red side ($\lambda > 6000$\,\AA) spectral features are measured from separate observations, and hence it is difficult to accurately constrain the relative flux calibration between the two sides.}
		\begin{tabular}{ccccc}
			\hline
			\hline
			Line & Wavelength (\AA) & Flux ($10^{-16}$\,erg\,cm$^{-2}$\,s$^{-1}$) &  $\sigma_V$ (km\,s$^{-1}$) & $EW$ (\AA)  \\
			\hline
			H$\delta^\dagger$ & $4103$ & - & - & $2.4 \pm 0.7$ \\
			H$\beta$ & $4861$ & $68 \pm 5.7$ & $178 \pm 18$ & $2.3 \pm 0.3$\\
			\lbrack O III\rbrack & $5007$ & $25 \pm 11$ & $290 \pm 180$ & $0.9 \pm 0.4$ \\
			Na D$^\dagger$  & $5894$ & - & - & $2.4 \pm 0.4$ \\
			\lbrack N II\rbrack & $6548$ & $1125 \pm 74$ & $199 \pm 15$ & $2.2 \pm 0.1$ \\
			H$\alpha$  & $6563$ & $2654 \pm 520$ & $156 \pm 3$ & $5.2 \pm 0.1$\\
			\lbrack N II\rbrack & $6584$ & $3111 \pm 63$ & $180 \pm 4$ & $6.1 \pm 0.2$ \\
			\lbrack S II\rbrack & $6717$ & $1384 \pm 76$  & $209 \pm 13$ & $2.8 \pm 0.2$\\
			\lbrack S II\rbrack & $6731$ & $1015 \pm 51$ & $172 \pm 9$ & $2.0 \pm 0.1$\\
			He I & $10830$ & $9.0 \pm 0.5$ & $136 \pm 7$ & $2.9 \pm 0.2$ \\
			Pa\,$\gamma$ & $10942$ & $3.9 \pm 0.4$ & $106 \pm 13$ & $1.3 \pm 0.1$ \\
			Pa\,$\beta$ & $12822$ & $5.1 \pm 0.2$ & $73 \pm 3$ & $1.7 \pm 0.1$ \\
			Br\,$\gamma$ & $21661$ & $1.6 \pm 0.1$ & $93 \pm 8$ & $1.3 \pm 0.1$ \\
			\hline
		\end{tabular}
		
		\label{tab:host_spec}
	\end{table*}
	
	
	The source spectrum shows no clear evidence for broad spectral lines, as the estimated line widths are consistent with instrument resolution.
	We may infer the nature of the ionizing continuum responsible for the emission lines by computing the intensity ratios of different lines and placing these ratios on the frequently called BPT diagram \citep{1981PASP...93....5B}. In particular, we computed the following ratios: 
	log(\lbrack N II\rbrack6584 / H$\alpha$)  = $0.07 \pm 0.09$,  log(\lbrack O III\rbrack / H$\beta$)  = $-0.43 \pm 0.19$, and log(\lbrack S II\rbrack / H$\alpha$) = $-0.04 \pm 0.09$. These values point at a star forming region as the source of the emission lines \citep[e.g][]{2006MNRAS.372..961K}. In other words, the optical spectrum of NGC 7392 does not show evidence of an active galactic nucleus (AGN) hosted in this galaxy. This is further supported by the pre-flare WISE color of the galaxy, $W1-W2=0.044 \pm 0.08$, which is significantly lower than the typical AGN selection threshold $W1-W2 \geq 0.8$ \citep{2012ApJ...753...30S}.

	
	\section{Discussion} 
	\label{sec:disc}
	
	In this Letter, we report the identification of a remarkably bright mid-IR nuclear transient in the center of NGC 7392, which reached a $W2$ peak luminosity of $\nu L_\nu \simeq 10^{43} ~\text{erg}\,\text{s}^{-1}$. Based on the galaxy's pre-flare mid-IR color and the intensity ratio of optical lines, NGC 7392 was inferred to not host an AGN. Moreover, while NGC 7392 is a star forming galaxy, the large radiated energy output of the transient ($E_\text{rad,BB} > 3 \cdot 10^{51} ~\text{erg}$) exceeds significantly the typically observed output and the theoretical upper limit expected in the case of a supernova \citep[e.g.][]{2016ApJ...820L..38S}. 
	
	Superluminous supernovae (SLSNe) are rare stellar explosions that may feature such a large radiated energy from the deaths of extremely massive stars. However, SLSNe, with an event rate smaller than that of TDEs \citep[e.g.][]{2013MNRAS.431..912Q}, have been found to occur predominantly in low-metallicity galaxies \citep[e.g.][]{2019ARA&A..57..305G}, with galactic masses that are well below $\sim 10^{10} M_\odot$  \citep{2015MNRAS.449..917L}. The estimated stellar mass of NGC 7392 is an order of magnitude larger than this limit (Sect. \ref{sect:host_photo}). We also do not detect any evidence for broad nebular emission lines even in our NIR spectra (where the dust extinction is less effective) as seen in SLSNe at late phases \citep{Jerkstrand2017}. Together, we find it highly improbable that a SLSN event produced \wtp. 
	
	Instead, the temporal evolution of the transient's emission follows closely the flux evolution expected for the intrinsic emission of a TDE (i.e. $F \propto t^{-5/3}$, Fig. \ref{fig:bololc}). The observed mid-IR flare, though, corresponds to the dust echo of this intrinsic flux and its evolution depends on both the intrinsic emission and the dust distribution. Assuming for simplicity a boxcar response function for the dust reprocessing \citep[e.g.][]{2016ApJ...829...19V}, the observed mid-IR light curve will have a temporal evolution similar to that of the reprocessing emission if the width of the response function is comparable to the cadence of the IR light curve, which is the case for \wtp\ (Sect. \ref{sec:flare}). Therefore, and after excluding other potential explanations, we conclude that \wtp\ was most likely the result of a TDE.

	NGC 7392 is a nearby galaxy, located at around 42.3 Mpc. To the best of our knowledge, this is the closest TDE detected in the last decade. Yet, there was no flare detected in its optical emission. Physically, this was probably the result of heavy obscuration of the transient source, which made it undetectable in the optical waveband. Alternatively, the source may have been intrinsically faint in the visible spectrum, while it was still UV bright\footnote{It should be noted that given the distance of NGC 7392, the lack of an optical flare in the ASASSN data and the lack of detection of an X-ray transient event by MAXI strongly favor the scenario of an obscured TDE. For instance, even a moderate luminosity of $L_X \sim 2 \cdot 10^{43} ~\text{erg s}^{-1}$ would have made \wtp\ detectable by MAXI. }.  
	In both scenarios, the observed mid-IR emission is the result of dust reprocessing of the intrinsic UV/soft X-ray radiation of the TDE by a surrounding dusty medium. To explain the large mid-IR flux, the obscuring dusty medium probably has a considerable covering factor, significantly larger\footnote{A precise estimate of the dust covering factor is not possible since the transient was not detected in the optical band.} than 1\% that has been inferred for the obscuring medium of typical optically detected TDEs \citep{2016ApJ...829...19V, 2021ApJ...911...31J}. A similarly high covering factor, of around 20\%, was inferred for the case of ATLAS17jrp, a TDE with strong IR emission \citep{2022ApJ...930L...4W}. Additionally, our analysis of the observed mid-IR emission of \wtp\  suggests a medium located close to the central black hole, around $ 0.05-0.2~ \text{pc}$ away from it. 

	After the identification of \wtp, we obtained an optical and NIR spectrum of the source as part of our follow-up. The source spectrum exhibits no broad spectral features that are usually associated with TDEs \citep[e.g.][]{2021ApJ...908....4V}. While this could perhaps be explained if \wtp\ is part of the newly introduced TDE-featureless class \citep{2022arXiv220301461H}, it is most likely a consequence of the central source ionizing continuum being currently very dim. Moreover, the source was not formally detected in recent X-ray observations, potentially due to significant absorption along our line of sight or due to the source being currently intrinsically faint in X-rays. Finally, the non-detection of the source in VLASS constraints its radio luminosity at $3$\,GHz to $\lesssim 5 \times 10^{36}$\,erg\,s$^{-1}$, making it fainter than nearly all radio-detected TDEs at similar phases \citep{Alexander2020}.
	
	Using the well studied $M-\sigma$ relation, we were able to get a rough estimate for the mass of the central black hole, $\log (\frac{M_{BH}}{M_\odot}) = 7.5^{+0.2}_{-0.3}$, which lies on the higher end of previously detected TDEs \citep[e.g.][]{2021ARA&A..59...21G, 2022arXiv220301461H}. For the estimated black hole mass, the corresponding Eddington limit is $L_\text{Edd} \simeq 4 \cdot 10^{45} ~\text{erg s}^{-1}$. While TDEs are known to emit close to the Eddington limit \citep[e.g.][and references therein]{2021ARA&A..59...21G}, the above value is significantly higher than the maximum mid-IR luminosity of \wtp\ (Fig. \ref{fig:bololc}), indicating that the observed dust reprocessed emission might correspond only to a fraction of the source bolometric luminosity as has also been demonstrated in theoretical explorations \citep{2016MNRAS.458..575L}.
	
	In addition, it is interesting to note again that while the observed TDE \wtp\ is one of the closest TDEs ever observed, it was missed by optical searches. This hints at the existence of a TDE population that have gone unnoticed by traditional methods so far, as was recently suggested by \cite{2021ApJS..252...32J}.

	Further, in contrast to most known TDEs which occupy the green valley (Fig.~\ref{fig:urmgal}), \wtp\ took place in the center of a star forming galaxy. NGC 7392 has been deduced to currently have a moderate star forming rate of $\log (\frac{\text{SFR}}{M_\odot \cdot \text{yr}^{-1}}) \simeq 0.7-0.8$ \citep{Iglesias2006}. The star-forming nature of NGC 7392 is further confirmed by its position in the $u-r$ vs. $M_\text{gal}$ plane (Fig. \ref{fig:urmgal}). This is in stark contrast to optically and X-ray detected TDEs, which are preferentially found in particular post-starburst and green valley galaxies \citep[e.g.][]{2016ApJ...818L..21F, 2021ApJ...908....4V, 2021MNRAS.508.3820S}. Consequently, our results not only suggest a population of TDE missed by traditional surveys, but in addition, this population may have a different distribution of host galaxies. Our analysis supports the framework put forth by \cite{2021ApJ...910...93R}, who suggested that dust obscuration suppresses the detection of TDEs in star-forming galaxies, although these are expected to be similarly common to known TDEs. \cite{2022ApJ...930L...4W} have shown that ATLAS17jrp, a TDE with a high mid-IR luminosity, took place in a star forming galaxy as well, which further supports that mid-IR bright TDEs may be the key in obtaining a complete sample of the TDE population.
	
	The identification of a very bright TDE dust reprocessing signal with no optical flaring counterpart in a nearby galaxy highlights that mid-IR surveys offer a unique opportunity to detect and probe this historically overlooked TDE population. Yet, its archival identification more than 7 years after the outburst onset highlights the potential for real-time identification of such events utilizing both real-time image subtraction pipelines (which was unavailable for the NEOWISE mission) as well as faster cadence access to survey data (noting that NEOWISE data is publicly released only once a year). In addition, upcoming ground-based and higher cadence NIR surveys such as the Wide Field Infrared Transient Explorer \citep{Lourie2020} and the Prime Focus Infrared Microlensing Experiment\footnote{\url{http://www-ir.ess.sci.osaka-u.ac.jp/prime/index.html}} may provide new avenues to finding similar events. Similar ground-based searches targeting Ultra-luminous Infrared Galaxies \citep{2018Sci...361..482M, 2020MNRAS.498.2167K} have indeed been more fruitful in real-time identification and spectroscopic follow-up for detecting spectroscopic accretion signatures, while the earliest follow-up for \wtp\ was obtained after the delayed identification. As purely mid-IR emission dominated phenomena, the James Webb Space Telescope offers the only promising opportunity to study sources like \wtp\ at the peak of the dust emission SED in order to reveal its temperature and composition, and thereby the bolometric luminosity and Eddington fraction, as well as the mid-IR fine structure and molecular lines to constrain the central ionizing source.

	\section*{Acknowledgements}
	We thank D. Dong and M. MacLeod for valuable discussions. We also would like to thank the \nustar\ and \swift\ teams for the approval of our Director's Discretionary Time requests and for carrying out the observations. K. D. was supported by NASA through the NASA Hubble Fellowship grant \#HST-HF2-51477.001 awarded by the Space Telescope Science Institute, which is operated by the Association of Universities for Research in Astronomy, Inc., for NASA, under contract NAS5-26555. This publication makes use of data products from the Wide-field Infrared Survey Explorer, which is a joint project of the University of California, Los Angeles, and the Jet Propulsion Laboratory/California Institute of Technology, funded by the National Aeronautics and Space Administration. This paper includes data gathered with the 6.5 meter Magellan Telescopes located at Las Campanas Observatory, Chile. Based on observations obtained at the Southern Astrophysical Research (SOAR) telescope, which is a joint project of the Minist\'{e}rio da Ci\^{e}ncia, Tecnologia e Inova\c{c}\~{o}es (MCTI/LNA) do Brasil, the US National Science Foundation’s NOIRLab, the University of North Carolina at Chapel Hill (UNC), and Michigan State University (MSU). 
	
	\bibliographystyle{aasjournal}	
	\bibliography{wtp14_arxiv}{}
	
\end{document}